# Narrowband-to-broadband switchable and polarization-insensitive terahertz metasurface absorber enabled by phase-change material

S. Hadi Badri[1], M. M. Gilarlue[1], Sanam SaeidNahaei[2,*], Jong Su Kim[2,*]

[1] Department of Electrical Engineering, Sarab Branch, Islamic Azad University, Sarab, Iran.

[2] Department of Physics, Yeungnam University, Gyeongsan, 38541, Republic of Korea.

[*] sanam.nahaei@yu.ac.kr (S. SaeidNahaei) and jongsukim@ynu.ac.kr (J. S. Kim)

## Abstract

A terahertz absorber with controllable and switchable bandwidth and insensitive to polarization is of great interest. Here, we propose and demonstrate a metasurface absorber with switchable bandwidth based on a phase-change material of vanadium dioxide ($VO_2$) and verify its performance by the finite element method simulations. The metasurface absorber is composed of a hybrid cross fractal as a resonator separated from a gold ground-plane by a polyimide spacer. Switching from narrowband to broadband absorber is achieved via connecting $VO_2$ patches to the gold first-order cross fractal converting the resonator to a third-order cross fractal. In the insulator phase of $VO_2$, the main narrowband absorption occurs at the frequency of 6.05 THz with a 0.99 absorption and a full-width half-maximum (FWHM) of 0.35 THz. Upon insulator-to-metal transition of $VO_2$, the metasurface achieves a broadband absorption with the FWHM of 6.17 THz. The simulations indicate that by controlling the partial phase-transition of $VO_2$, we can tune the bandwidth and absorption level of the absorber. Moreover, the designed absorber is insensitive to polarization due to symmetry and works well for a very wide range of incident angles. In the metallic state of $VO_2$, the absorber has an absorption exceeding 0.5 in the 3.57-8.45 THz frequency range with incident angles up to 65°.

## 1. Introduction

Manipulating electromagnetic waves with metamaterials [1-4], metasurfaces [5, 6], plasmonic structures [7, 8], nanoparticles [9], and transformation optics [10-13] have attracted great interest. Metamaterials are artificially engineered composite materials with unique properties not found in natural materials. Metasurfaces are the two-dimensional counterparts of metamaterials with subwavelength thickness and they have attracted increasing attention due to their planar geometry and relative ease of fabrication. The metasurfaces rely on local modification of the boundary conditions for an incident electromagnetic wave while conventional optical components manipulate the wave by gradual phase accumulation along the curved and bulky components. The novel arrangement of subwavelength scatterers in the metasurfaces has enabled us to engineer the amplitude and phase of the scattered electromagnetic wave in the densely integrated and miniaturized devices [14]. Various metasurfaces with unique and fascinating functionalities such as holograms [15, 16], polarization conversion [17, 18], antennas [19, 20], lenses [21, 22], beam steering [23], invisibility cloaks [24], electromagnetically induced transparency (EIT) [25], and absorbers [26] have been introduced. However, most of the reported metasurfaces offer a fixed functionality. Following the developments in the integrated circuits, components with switchable and multiple functionalities are essential in paving the path to a miniaturized device with multiple functionalities.

Recently, interesting devices such as on-chip integrated photonics components [27, 28] and memories [29, 30] have been introduced based on phase-change materials (PCM). Various PCMs have been utilized to design metasurfaces. The two prominent PCMs that have been studied are $Ge_2Sb_2Te_5$ (GST) and vanadium dioxide ($VO_2$). Dynamic thermal emission control in a metasurface has been proposed based on GST. In this study, GST is used as a spacer in the metal-GST-metal configuration [31]. The imaging function of a Fano-resonant metasurface assisted by GST has also been studied. In this metasurface, the shift of resonance when GST film on the nanodisk transitions from amorphous to crystalline state manipulates the transmitted power at a specific wavelength [32]. Active optical radiation manipulation of GST-based metasurfaces has been studied. A metasurface with silicon double-nanodisks on a GST spacer can dynamically control the optical radiation strength and spectral position of the scattering minimum [33]. Metasurfaces based on GST with various tunable functionalities have been described in some review articles [34]. However, $VO_2$ offers higher conductivity change upon phase transition in the terahertz (THz) range compared to GST. Therefore, unique THz metasurfaces with switchable functionalities have been proposed based on $VO_2$. $VO_2$ is a phase-change material with reversibly switchable insulator and metallic phases controlled through external stimulus such as heat. In the THz range, the difference between the electric conductivity in the insulator and metallic phases of $VO_2$ is about three orders of magnitude making it possible to design novel devices. A variety of switchable and diversified functionalities in a single metasurface has been proposed such as a metasurface that supports broadband EIT with a bandwidth of 0.27 THz and a maximum transmission reaching 0.83 when $VO_2$ is in its insulating phase. Upon phase-transition, it operates as a broadband absorber with the total absorption exceeding 0.93 in a bandwidth of 0.5 THz [35]. A bifunctional metasurface has been presented that can be switched from a broadband absorber to a reflective broadband linear polarization converter. When $VO_2$ is in the metallic state, high absorptance in the 0.52-1.2 THz range is achieved while in the insulator state, the metasurface becomes a broadband linear polarization converter with the reflectance of 0.9 in the 0.42-1.04 THz range [36]. A reconfigurable metasurface composed of a $VO_2$ square ring, polyimide spacer, and $VO_2$ film has been designed with broadband absorption and linear-to-circular polarization conversion functionalities. In the insulator phase, the metasurface is a linear-to-circular polarization converter with ellipticity close to -1 and axis ratio $<3$ dB from 1.47 THz to 2.27 THz. Upon phase-transition of $VO_2$, the absorption is above 0.9 from 0.74 THz to 1.62 THz [37]. Tunable coding metasurfaces have been proposed based on C-shaped aluminum-$VO_2$ hybrid resonators switched from anisotropic structures to quasi-isotropic ones due to the insulator-to-metal transition of $VO_2$. The designed metasurfaces realize the simultaneous high-efficiency modulation of the polarization and wavefront of THz beams, including linear and circularly polarized ones [38]. A tunable dual broadband absorber has been designed with absorption higher than 0.8 in the 0.56-1.44 THz and 2.88-3.65 THz frequency ranges. The absorption level can be adjusted from 0.2 to 0.9 through the operating temperature of $VO_2$ [39]. An active absorber has been proposed based on a silicon nanograting embedded with $VO_2$ placed on a silver mirror. Under the insulator state of $VO_2$, double narrowband perfect absorptions are achieved at the resonant wavelengths of 1077 and 1314 nm. Under the metallic $VO_2$ phase, the proposed design demonstrates triple adjacent perfect absorptions leading to a broadband absorption [40].

Most metasurface absorbers have a fixed spectral response or a single functionality. Therefore, tunable bi-functional metasurfaces are of great interest for THz applications. Recently, metasurfaces based on metal-$VO_2$ hybrid resonators have been proposed for THz applications such as wavefront and polarization manipulation [38], ultrafast modulation [41], and bandstop-to-bandpass converter [42]. While the combination of metal and $VO_2$ in the resonator structure enhances the modulation depth [40], the metasurfaces based on metal-$VO_2$ hybrid resonators have not been studied extensively for THz applications. In this paper, we design a tunable absorber based on $VO_2$ hybrid metasurface at THz frequencies. The metasurface consists of a cross fractal metal-$VO_2$ hybrid resonator placed on the polyimide spacer and a golden ground-plane. The designed resonator can be switched from a first-order cross fractal to a third-order cross fractal due to the insulator-to-metal transition of $VO_2$. This enables us to switch from a narrowband to a broadband absorption where the full-width half-maximum (FWHM) of the absorber is broadened by about fifteen times. When $VO_2$ is in the insulator phase, the resonator transforms to the first-order cross fractal with a narrowband absorption at the resonant frequencies of 3.97, 6.05, and 8.11 THz with absorption exceeding 0.5 in the bandwidth of 0.38, 0.35, 0.22 THz, respectively. When $VO_2$ undergoes the insulator-to-metal transition (IMT), the absorption bandwidth increases to the FWHM of 6.17 THz centered around 5.79 THz. We also examine the performance of the absorber in the case of partial phase-transition of $VO_2$ and oblique incidence. The designed switchable and polarization-insensitive metasurface absorber has the advantages of multi-functionality and a simple configuration compared with multilayer metamaterials such as [43]. Furthermore, dynamic thermal tuning of the absorption spectrum can provide diverse applications. When $VO_2$ is in the metallic state, the metasurface's broadband absorption can be used in anti-reflection coatings, thermal imaging systems, cloaking, and photovoltaic devices [44, 45]. It is also useful for broadband vector excitation control and broadband vector pulse shaping [46, 47]. In the insulator phase, the narrowband absorption of the absorber can be utilized in switches, filters, modulators, sensors, detectors, and coherent thermal emitters [40, 48].

## 2. Design of the switchable metasurface

The schematic of the proposed metasurface is illustrated in Fig. 1. The metasurface is composed of a polyimide (PI) spacer with a thickness of $t_s$=8 μm placed on a gold ground-plane with a thickness of $t_{Au}$=200 nm. On top of the polyimide spacer. a first-order cross resonator made of gold (Au) with a length of $L_1$=20 μm is placed at the center of the unit cell. The added $VO_2$ strips, with a thickness of $t_{VO_2} = 200\ nm$, convert the resonator to a third-order cross fractal. The optimized lengths of the $VO_2$ strips are $L_2$=5 µm, $L_3$=4.5 µm, $L_4$=6 µm. The unit cell is periodically arranged in both the x- and y-directions with a pitch of $P$=40 µm. The reason for choosing the cross fractal as the resonator is its symmetry which provides a polarization-insensitive performance, i.e., its spectral response is the same for the incident $x$- or $y$-polarized electromagnetic wave. The conductivity of $VO_2$ in its metallic phase is in the order of $10^5$ S/m which is considerably lower than the conductivity of gold ($10^7$ S/m). Hence, the modulation depth is limited when only $VO_2$ is used as a resonator. The combination of metal and $VO_2$ in the resonator enhances the modulation depth [38]. When the $VO_2$ strips are in the insulator phase, the resonator reduces to a first-order cross fractal composed of gold strips providing narrow absorption bandwidth. Upon phase-transition, the $VO_2$ strips' conductivity increases and $VO_2$ strips contribute to the absorption spectrum by converting the resonator to a third-order cross fractal. Moreover, $VO_2$ has a lower conductivity compared to gold increasing the loss in the resonator and, consequently, decrease the quality factor of the resonances which in turn broadens the absorption bandwidth of the absorber [14, 49]. The combination of gold and $VO_2$ in the resonator enables us to modulate the bandwidth and absorption level of the metasurface by controlling its temperature.

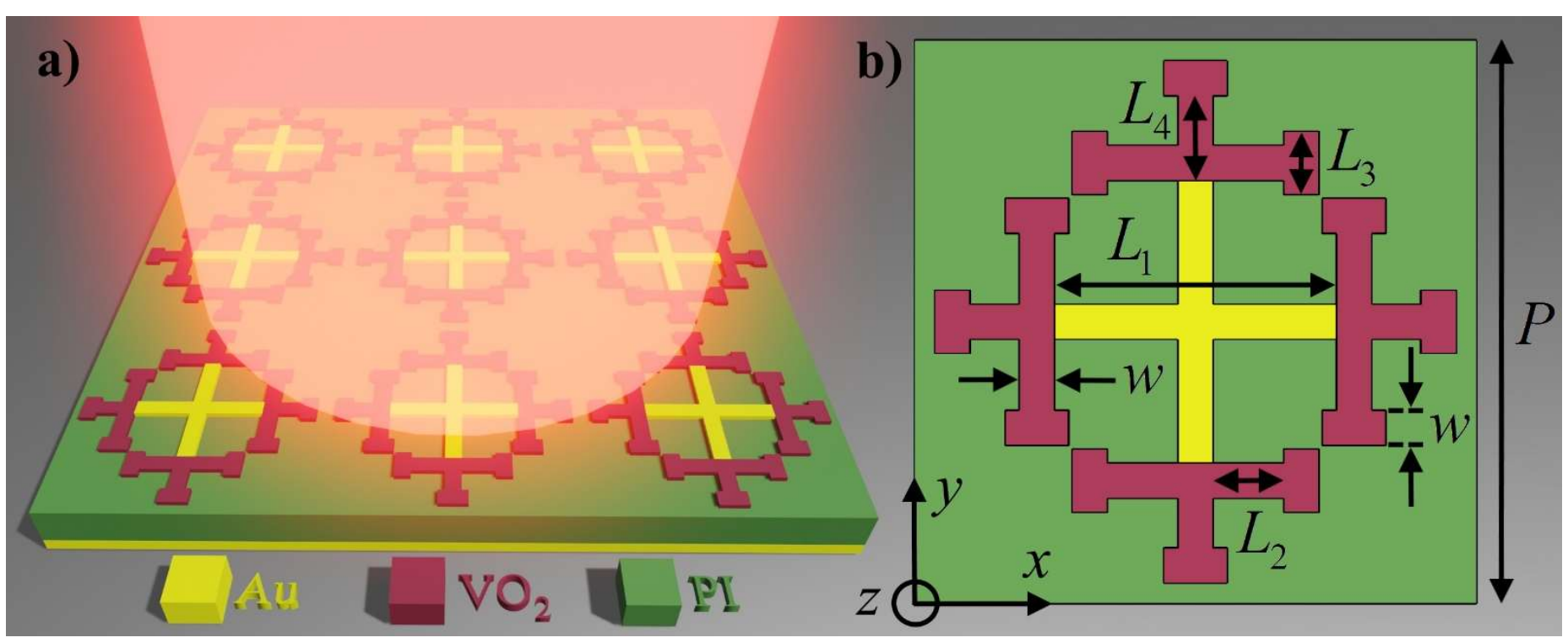

Fig. 1. a) The designed switchable metasurface absorber. b) Top view of the unit cell of the metasurface. The optimized geometrical parameters are $P$=40 µm, $L_1$=20 µm, $L_2$=5 µm, $L_3$=4.5 µm, $L_4$=6 µm, $w$=2.5 µm, $t_s$=8 µm, and $t_{Au} = t_{VO_2} = 200$ nm.

## 3. Results and Discussion

We use commercial CST Microwave Studio software for evaluating the performance of the designed switchable metasurface absorber through the finite element method (FEM). Periodic boundary conditions are applied along the $x$- and $y$-axes while a plane wave is normally incident along the $z$-axis. Since the resonator is symmetric, the performance of the metasurface for excitation with an electric field along the $x$- or $y$-axes are the same. The conservation of energy is utilized to calculate the absorption of the metasurface, i.e., $A(\omega)=1-T(\omega)-R(\omega)$. Considering that the thickness of the metallic ground-plane is much thicker than the skin depth, the transmission through the metasurface is $T(\omega)=0$. Therefore, the absorption is $A(\omega)=1- R(\omega)=1-|S_{11}|^2$ [50]. In the simulations, the relative permittivity of the lossy polyimide is $\varepsilon_{PI}=2.88+0.09i$ [51]. The

conductivity of gold is described by Drude model $\sigma_{Au}=\varepsilon_0\omega_p^2/(\gamma-i\omega)$, where $\varepsilon_0$ is the dielectric constant in vacuum, $\omega$ is the angular frequency, $\omega_p=1.37\times10^{16}$ rad/s is the plasma frequency, and $\gamma=1.2\times10^{14}$ rad/s represents the collision frequency [50, 52]. At room temperature, $VO_2$ is in its insulator state with a permittivity of $\varepsilon_i=9$ and a conductivity of $\sigma_i=200$ S/m. When the temperature is beyond 85 °C, $VO_2$ switches to its metallic state with a conductivity of $\sigma_m=2\times10^5$ S/m [14, 35]. The absorption spectra of the absorber for $VO_2$ strips in the insulator and metallic phases are displayed in Fig. 2 for a structure with the optimized geometrical parameters of $P$=40 µm, $L_1$=20 µm, $L_2$=5 µm, $L_3$=4.5 µm, $L_4$=6 µm, $w$=2.5 µm, $t_s$=8 µm, and $t_{Au}=t_{VO_2}=200$ nm. The electric field intensity of the resonant modes at the interface of the spacer and the resonator and at the frequencies of 3.97, 6.05, and 8.11 THz are calculated and plotted for the insulator and metallic phases of $VO_2$ in Fig. 3. In the insulator phase, the electric field concentrates around the gold strips while after phase-transition the electric field intensity is higher around the $VO_2$ strips.

A possible method for fabricating the proposed metasurface is the micromachining process which includes the following steps [42]. First, a 200 nm thick $VO_2$ layer is deposited on the polyimide spacer by reactive RF magnetron sputtering followed by annealing. Then a photoresist is spin-coated on the $VO_2$ film and $VO_2$ film is patterned by ion beam etching. Finally, the 200 nm thick structured gold film is formed using the lift-off process. To control the operating temperature of the proposed absorber, the electric current cannot be applied since the gold strips are not connected to each other. When the metal strips are connected, the electric current passing through the metal strips can be used to tune the operating temperature [42, 53, 54]. Therefore, an external heater should be used [55, 56].

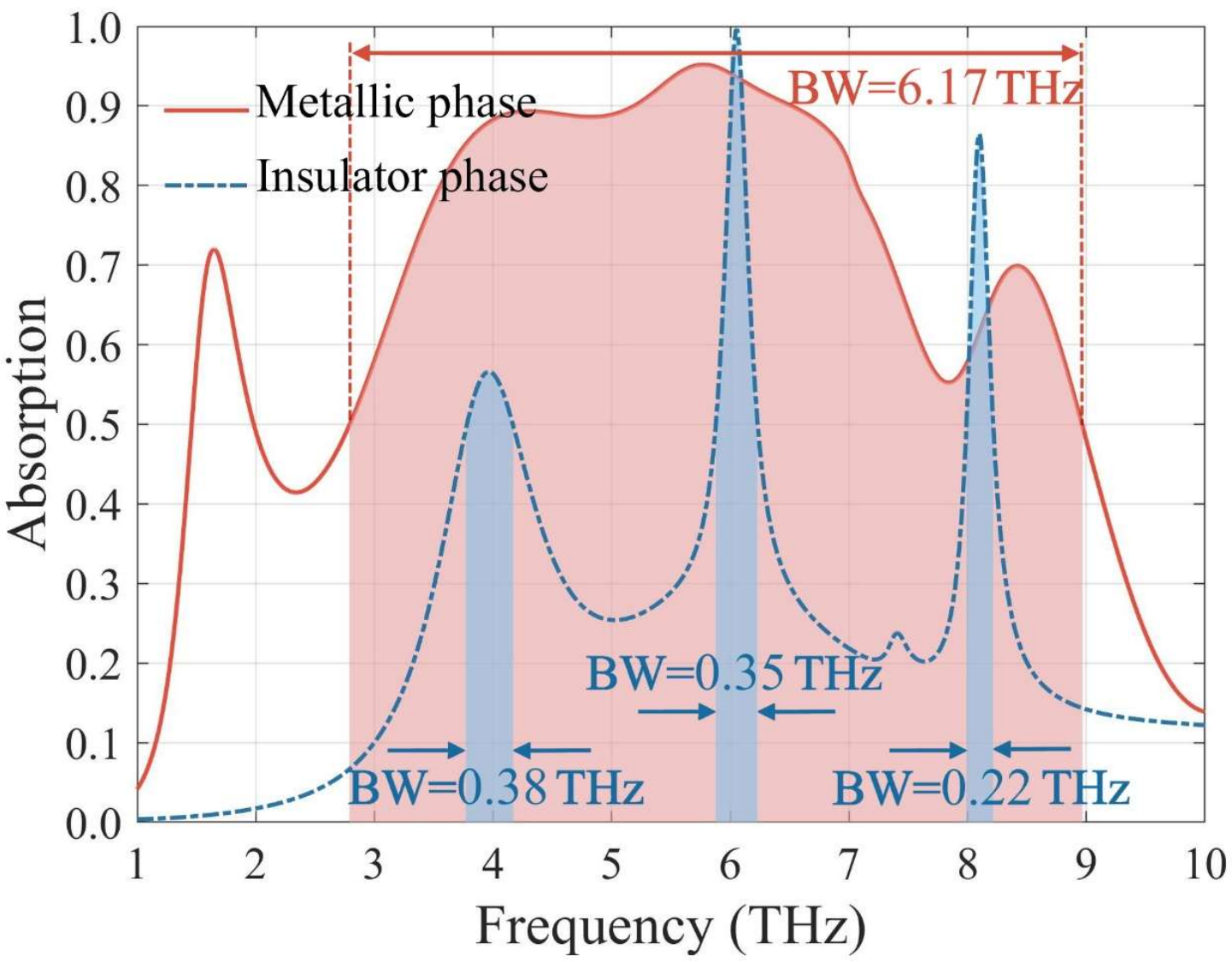

Fig. 2. Absorption spectra of the designed switchable metasurface corresponding to insulator and metallic phases of $VO_2$ at the normal incidence of the THz plane wave. The optimized geometrical parameters are $P$=40 µm, $L_1$=20 µm, $L_2$=5 µm, $L_3$=4.5 µm, $L_4$=6 µm, $w$=2.5 µm, $t_s$=8 µm, and $t_{Au}=t_{VO_2}=200$ nm.

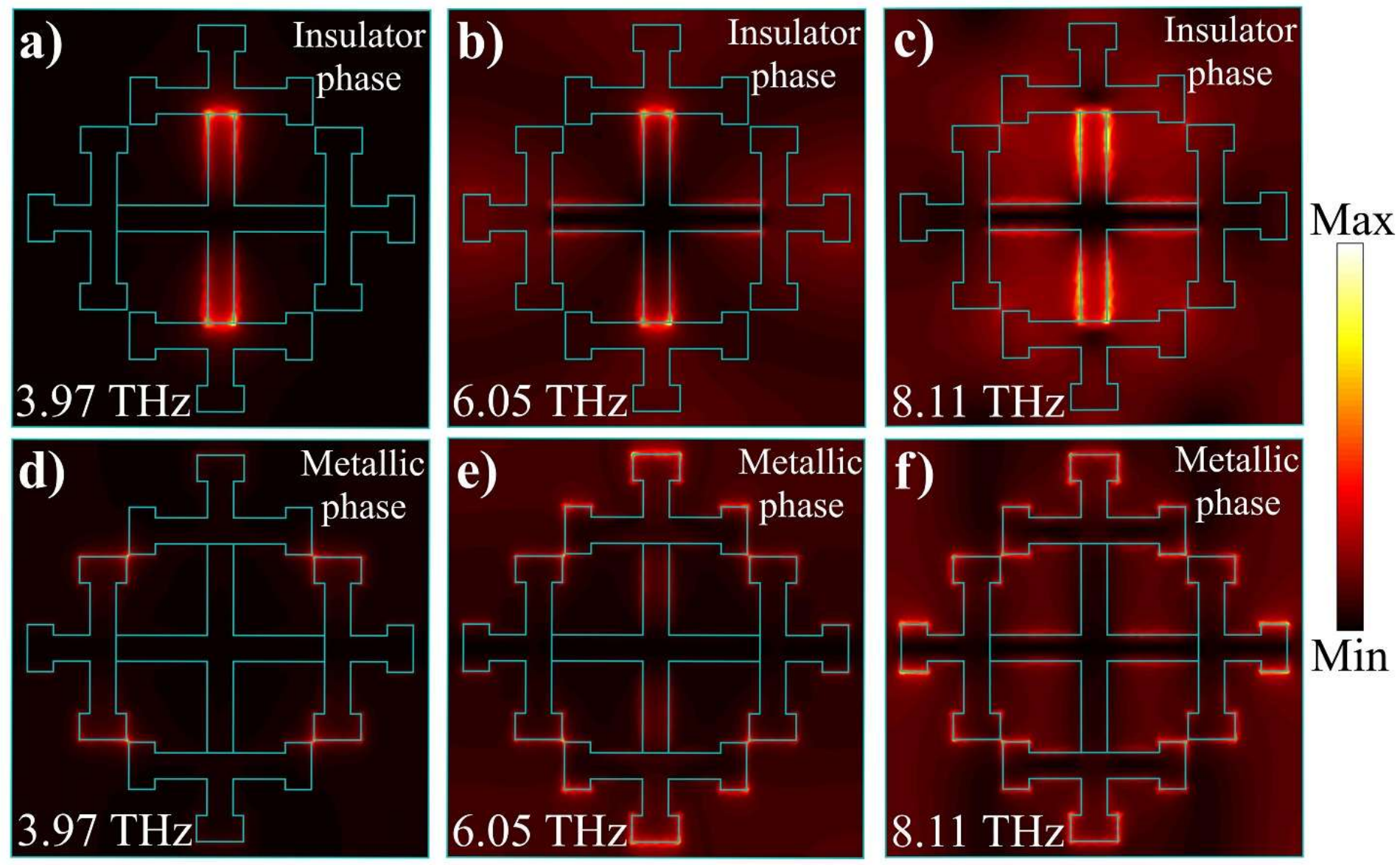

Fig. 3. Electric field distribution while $VO_2$ is in the insulator state at the frequency of a) 3.97, b) 6.05, and c) 8.11 THz at the interface of hybrid resonator and dielectric spacer. Electric field distribution after the insulator-to-metal transition of $VO_2$ at the frequency of d) 3.97, e) 6.05, and f) 8.11 THz. The plane wave is normally incident on the metasurface. The optimized geometrical parameters are $P$=40 µm, $L_1$=20 µm, $L_2$=5 µm, $L_3$=4.5 µm, $L_4$=6 µm, $w$=2.5 µm, $t_s$=8 µm, and $t_{Au} = t_{VO_2} = 200$ nm.

### 3.1 Effect of geometrical parameters

In this subsection, we investigated the geometrical parameters' effect on the spectral response of the designed switchable metasurface. First, we examine the effect of the first-order resonator's length ($L_1$) in the performance of the absorber while the other parameters are fixed to $P$=40 µm, $L_2$=5 µm, $L_3$=4.5 µm, $L_4$=6 µm, $w$=2.5 µm, $t_s$=8 µm, and $t_{Au} = t_{VO_2} = 200$ nm. In Fig. 4, the calculated absorption curves for different values of the first-order resonator's length ($L_1$) in the insulator (25 °C) and the metallic (85 °C) phases are shown. In the insulator phase, the absorption peak of 0.97 occurs at the frequency of 6.28 THz for $L_1$=15.0 µm. For $L_1$=17.5 µm, the narrowband absorption occurs at 6.13 THz with the maximum absorption of about unity. Increasing the length of the resonator to $L_1$=20.0 µm shifts the resonant peak to 6.05 THz. As shown in Fig 4(a), increasing the length of the first-order cross resonator composed of gold strips results in the increase of the effective electric length of the resonator and, consequently, the absorption peaks red-shift to lower frequencies [57]. Fig. 4(b) shows the broadband absorption spectra when the temperature is high enough to trigger phase-transition of $VO_2$ strips to the metallic phase. In this phase, the absorption bandwidth increases in comparison to the insulator phase. For $L_1$=15.0 µm, the average absorption in the 1-10 THz bandwidth is around 0.58 while increasing the $L_1$ to 17.5 µm increases the average absorption to 0.61. The structure with $L_1$=20.0 µm yields a higher average absorption of 0.66 in the 1-10 THz frequency range. Considering the above discussions, we chose $L_1$=20.0 µm.

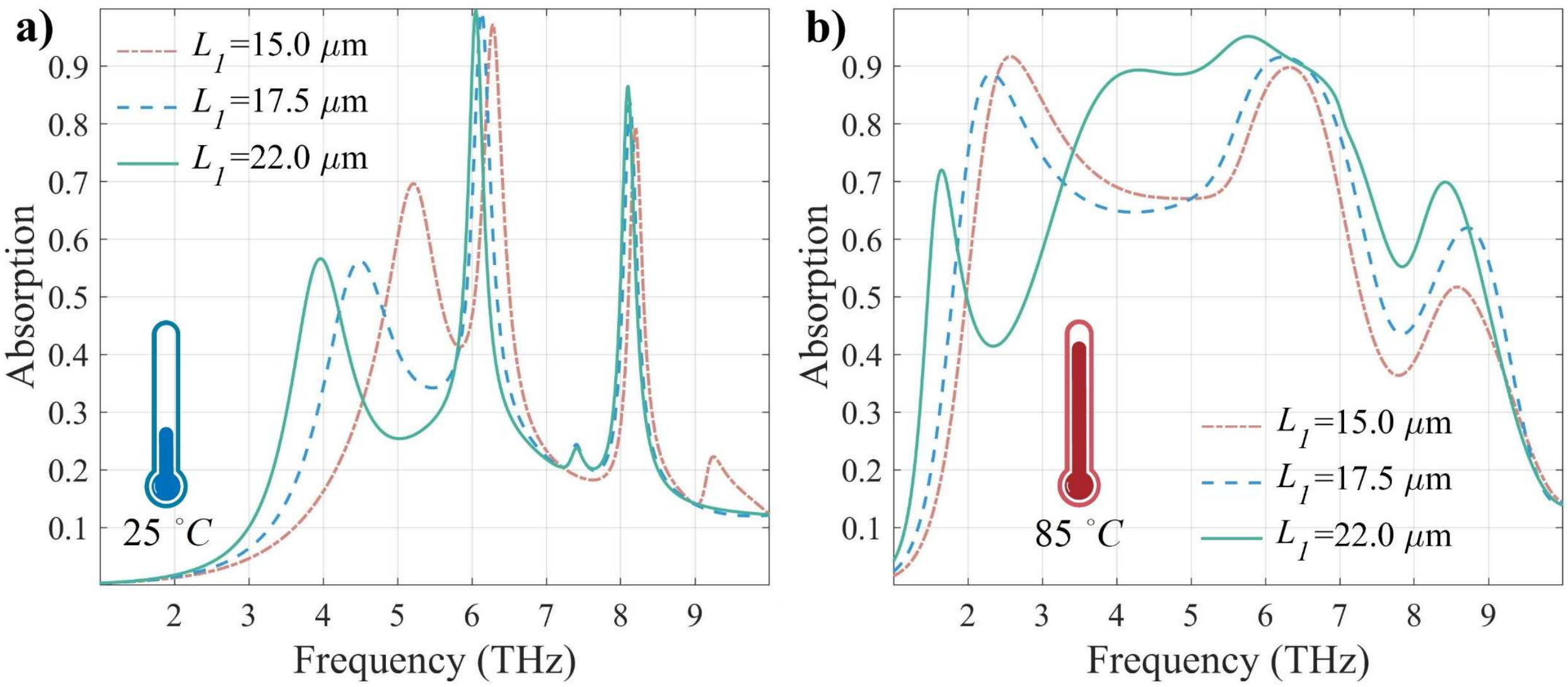

Fig. 4. The effect of first-order cross fractal's length ($L_1$) on the absorption spectra of the designed switchable metasurface corresponding to a) insulator and b) metallic phases of $VO_2$ at the normal incidence of the THz plane wave. The other geometrical parameters are $P$=40 µm, $L_2$=5 µm, $L_3$=4.5 µm, $L_4$=6 µm, $w$=2.5 µm, $t_s$=8 µm, and $t_{Au} = t_{VO_2} = 200$ nm.

The effect of the width of gold and $VO_2$ strips ($w$) on the absorption spectrum of the absorber is illustrated in Fig 5. In the insulator phase (Fig 5(a)), the absorption peaks and bandwidths do not show any significant dependency on the width of the gold and $VO_2$ strips. In this case, the resonator is effectively a first-order cross fractal composed of gold strips with very high conductivity, therefore, the width of the gold strips has a limited effect on the induced current and absorption performance of the metasurface. However, $VO_2$ strips in the metallic phase have about a hundred times lower conductivity compared to gold. Thus, increasing the width of $VO_2$ strips increases the effective conductivity of the resonator leading to stronger resonances with higher absorption levels. In the metallic phase and for $w$=1.0, 1.5, 2.0, and 2.5 µm, the average absorption in the 1-10 THz range is about 0.48, 0.56, 0.62, and 0.66, respectively. The structure with $w$=2.5 µm provides a higher average absorption, hence, it is chosen as a width of gold and $VO_2$ strips.

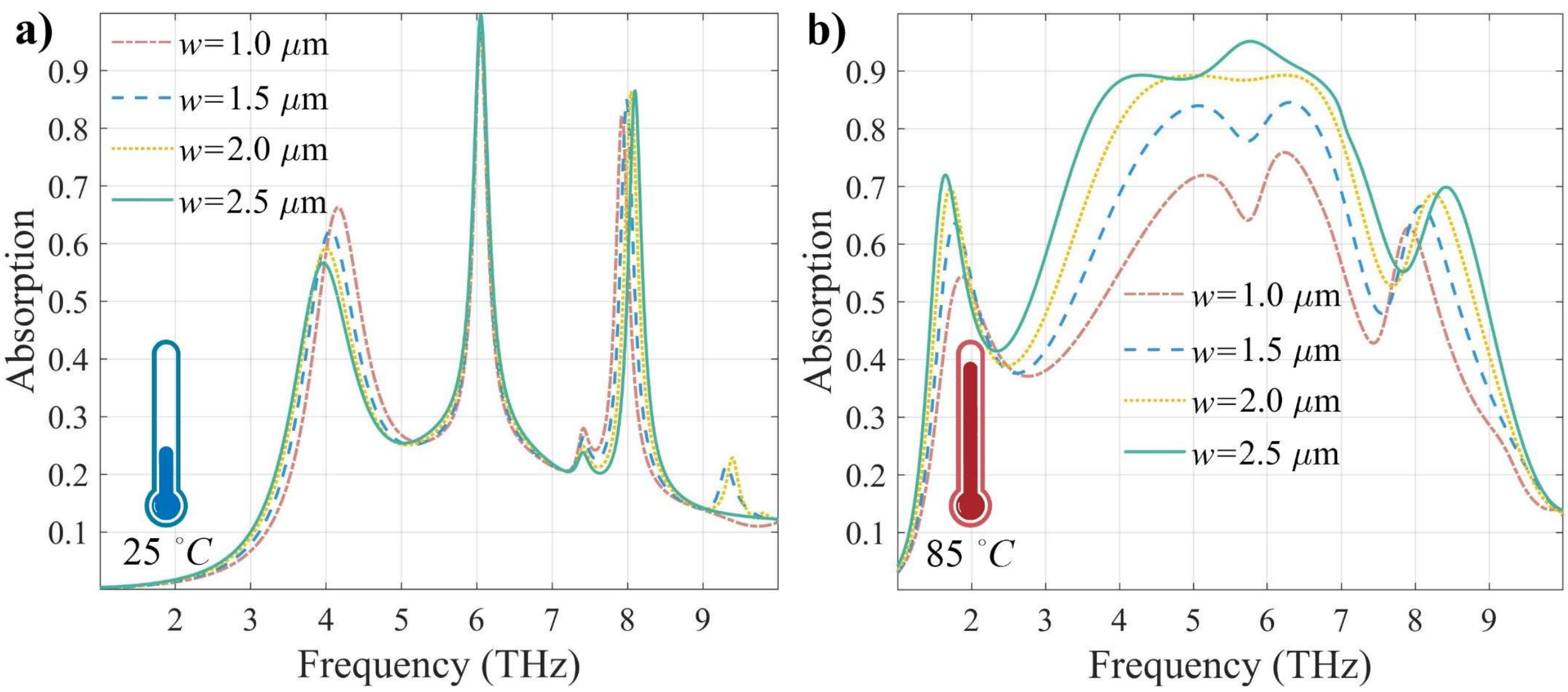

Fig. 5. The effect of gold and $VO_2$ strips' width ($w$) on the absorption spectra of the designed switchable metasurface corresponding to a) insulator and b) metallic phases of $VO_2$ at the normal incidence of the THz plane wave. The other geometrical parameters are $P$=40 µm, $L_1$=20 µm, $L_2$=5 µm, $L_3$=4.5 µm, $L_4$=6 µm, $w$=2.5 µm, and $t_{Au} = t_{VO_2} = 200$ nm.

The thickness of the dielectric spacer plays a critical role in adjusting the performance of the absorber. Therefore, we investigate the absorption spectrum for different values of spacer thickness ($t_s$). As the thickness of the spacer increases the electromagnetic coupling between the resonator and the ground-plane decreases leading to weaker resonances and, consequently, lower levels of absorption. As can be seen in Fig. 6(a), for $t_s$=6 µm, the average absorption level of the three absorption peaks is higher compared to thicker spacers due to a strong induced current loop between the resonator and the ground-plane [58]. Moreover, the absorption peaks red-shift to lower frequencies as the thickness of the spacer increases. Similarly, in the metallic state of $VO_2$, the center of the absorption band shifts to lower frequencies as the spacer's thickness increases. We chose $t_s$=8 µm so that the center of the FWHM absorption bandwidth in the metallic and insulator phases is about the same frequency, i.e., 6 THz. The main absorption peak in the insulator state is near 6 THz while in the metallic phase the center of bandwidth is about 5.68 THz.

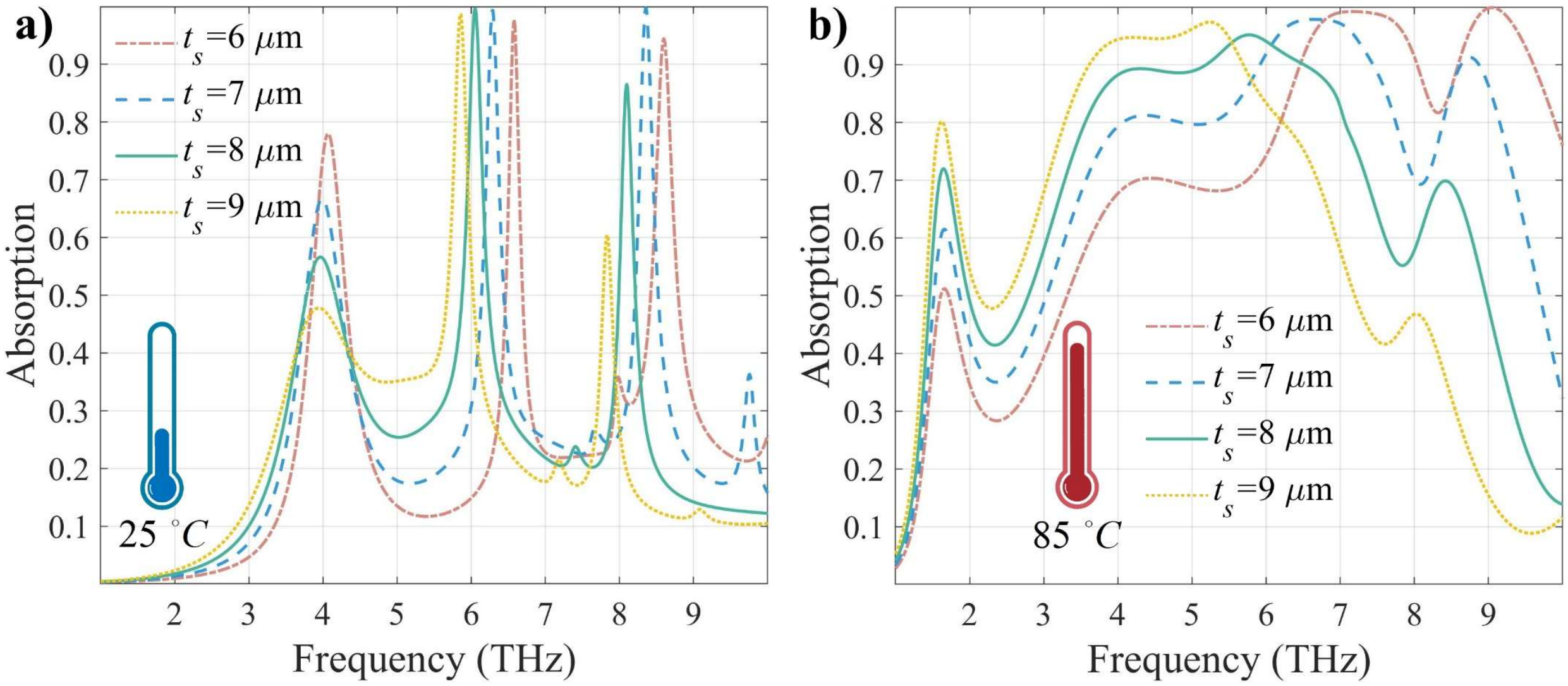

Fig. 6. The effect of spacer thickness ($t_s$) on the absorption spectra of the designed switchable metasurface corresponding to a) insulator and b) metallic phases of $VO_2$ at the normal incidence of the THz plane wave. The other geometrical parameters are $P$=40 μm, $L_1$=20 μm, $L_2$=5 μm, $L_3$=4.5 μm, $L_4$=6 μm, $w$=2.5 μm, and $t_{Au} = t_{VO_2} = 200$ nm.

The geometrical parameters determining the length and shape of $VO_2$ strips have no effect on the performance of the absorber in the insulator state. For instance, the effect of $L_4$ on the absorption spectra is shown in Fig. 7. As expected, the performance of the absorber is insensitive to the length and shape of $VO_2$ strips in the insulator state (Fig. 7(a)). However, in the metallic state, the maximum absorption level of the metasurface can be tuned by $L_4$. For $L_4$=4, 5, 6, and 7 μm, the maximum absorption level in the metallic state is 0.90, 0.93, 0.96, and 0.95, respectively. We chose $L_4$=6 μm since it provides a higher absorption level. We do not present the simulation results for $L_2$ and $L_3$ here, however, we should point out that similar to $L_4$, they only affect the performance of the absorber in the metallic state.

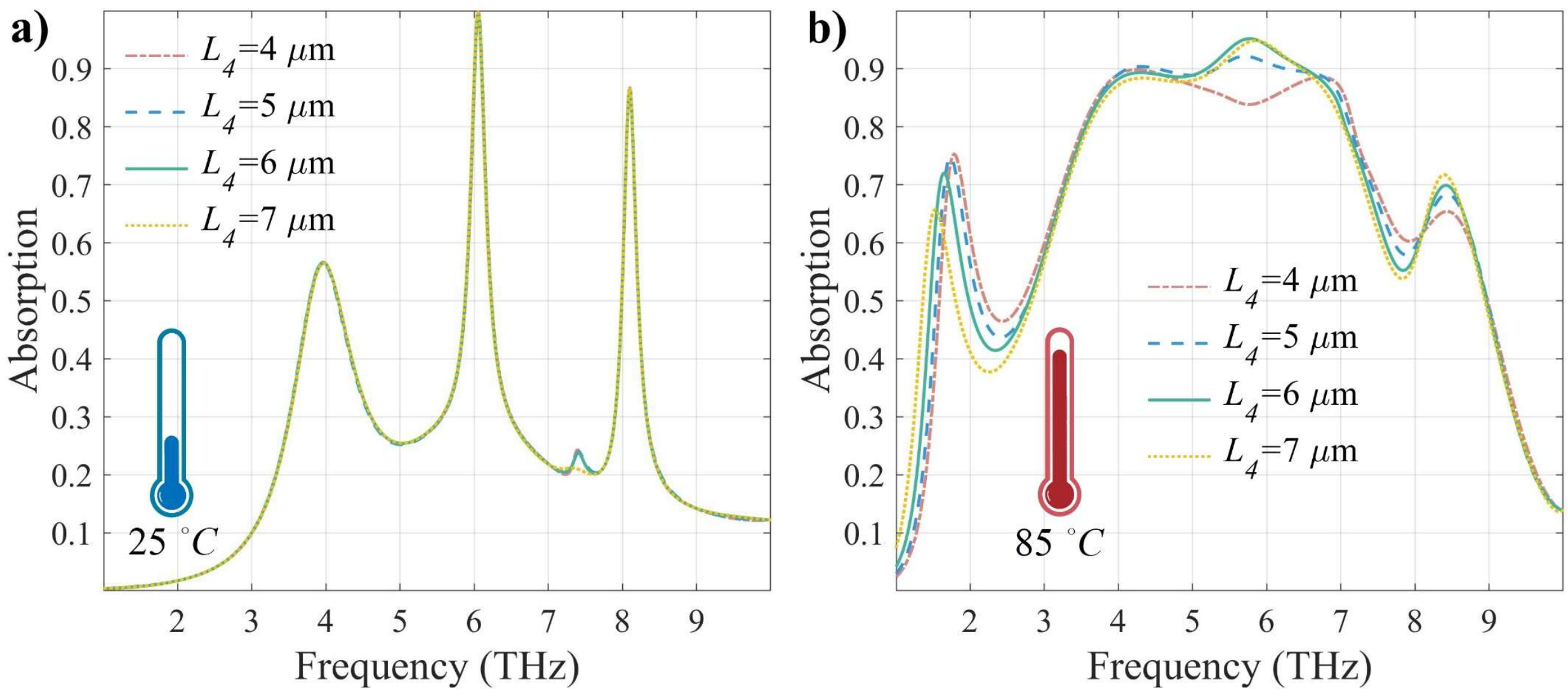

Fig. 7. The effect of $L_4$ on the absorption spectra of the designed switchable metasurface corresponding to a) insulator and b) metallic phases of $VO_2$ at the normal incidence of the THz plane wave. The other geometrical parameters are $P$=40 μm, $L_1$=20 μm, $L_2$=5 μm, $L_3$=4.5 μm, $w$=2.5 μm, $t_s$=8 μm, and $t_{Au} = t_{VO_2} = 200$ nm.

### 3.2 Partial phase-transition of $VO_2$

At room temperature, the $VO_2$ is in the insulator phase. As the temperature rises, the nucleation and growth of metallic puddles within the dielectric $VO_2$ is triggered. These metallic puddles expand and finally, the $VO_2$ is entirely in the metallic phase. The electrical and optical properties of the intermediate states, where both metallic and insulator states exist simultaneously, can be estimated by effective medium theory [59]. The effect of partial phase-transition of $VO_2$ on the performance of the absorber under normal incidence and while the conductivity changes from $2\times10^2$ to $2\times10^5$ S/m is illustrated in Fig. 8. The average absorption in the 1-10 THz range can be modulated from 0.27 to 0.66 by controlling the partial phase-transition of $VO_2$. For the conductivity of $2\times10^2$, $8\times10^3$, $2\times10^4$, $4\times10^4$, $1\times10^5$, and $2\times10^5$ S/m, the average absorption in the 1-10 THz range is 0.24, 0.45, 0.50, 0.53, 0.60, and 0.66, respectively.

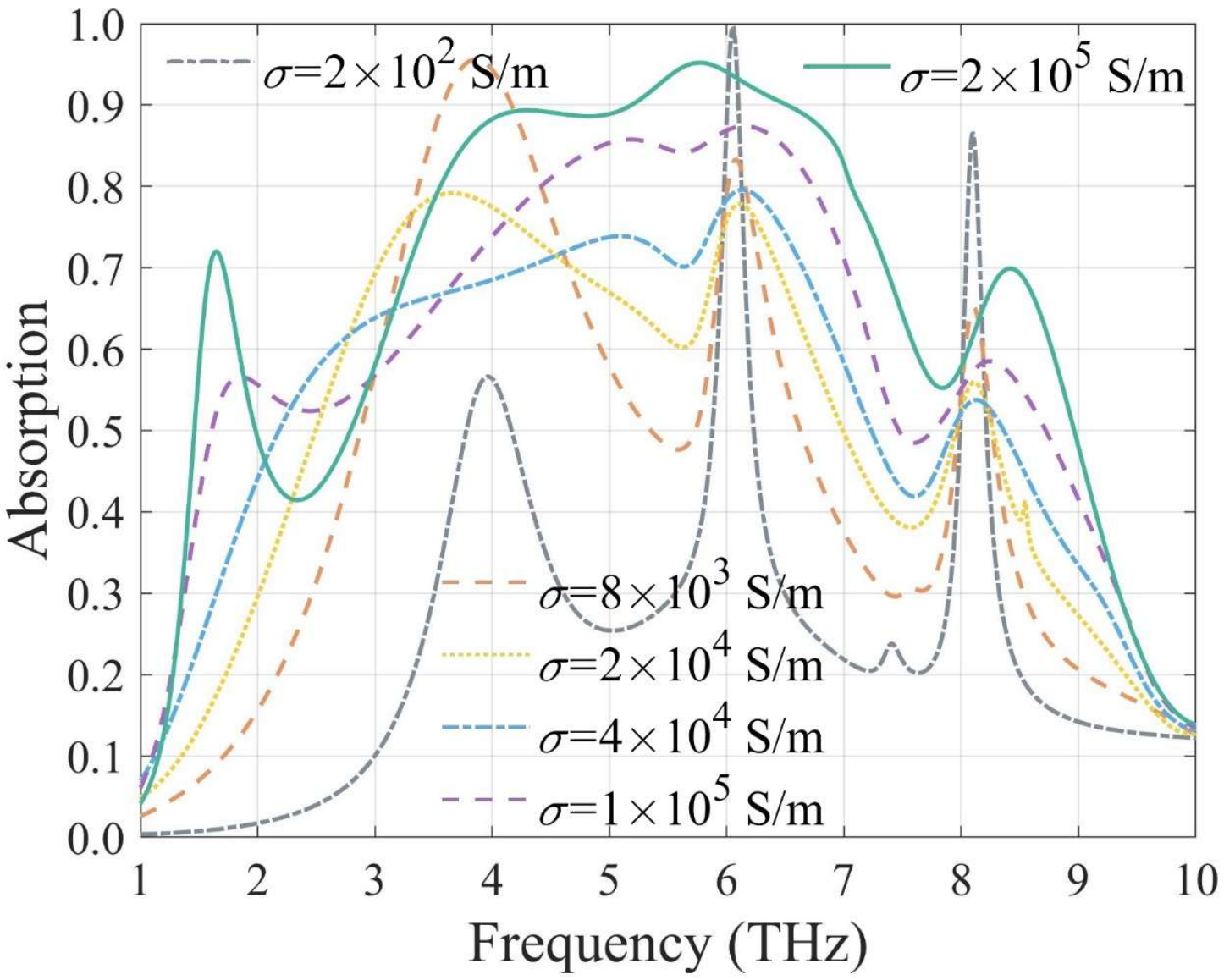

Fig. 8. Absorption spectra of the proposed metasurface with respect to the conductivity of $VO_2$ at the normal incidence of the THz plane wave. As the conductivity increases, the metasurface switches from a narrowband absorber to a broadband absorber. The optimized geometrical parameters are $P$=40 μm, $L_1$=20 μm, $L_2$=5 μm, $L_3$=4.5 μm, $L_4$=6 μm, $w$=2.5 μm, $t_s$=8 μm, and $t_{Au} = t_{VO_2} = 200$ nm.

The absorption bandwidth of the proposed metasurface absorber can also be interpreted by impedance matching, where the absorption approaches unity as the effective impedances of the absorber and the free space match. The absorption of the metamaterial based on the impedance matching theory and under normal incidence is [50]

$$A(\omega) = 1 - R(\omega) = 1 - \left|\frac{Z(\omega) - Z_0}{Z(\omega) + Z_0}\right|^2 = 1 - \left|\frac{Z_r(\omega) - 1}{Z_r(\omega) + 1}\right|^2 \tag{1}$$

where $Z(\omega)=\sqrt{\mu(\omega)/\varepsilon(\omega)}$ and $Z_0=\sqrt{\mu_0/\varepsilon_0}$ are the effective impedances of the absorber and the free space, respectively. The relative impedance between the absorber and the free space is $Z_r(\omega)=Z(\omega)/Z_0$. The scattering parameters retrieval technique can be utilized to characterize the metamaterial and calculate the relative impedance [60]

$$Z_r(\omega)=\sqrt{\frac{(1+S_{11}(\omega))^2-S_{21}(\omega)^2}{(1-S_{11}(\omega))^2-S_{21}(\omega)^2}} \tag{2}$$

$S_{11}$ and $S_{21}$ are scattering parameters where the first subscript denotes the receiving port and the second subscript denotes the excitation port. When the impedance of the absorber matches with that of the free space or equivalently the relative impedance equals unity ($Z_r$=1), reflection is minimized and absorption reaches near unity. The real and imaginary parts of the relative impedance are shown in Fig. 9. As the conductivity of $VO_2$ increases, the real part of $Z_r$ approaches one while the imaginary part of $Z_r$ approaches zero leading to broader absorption bandwidth.

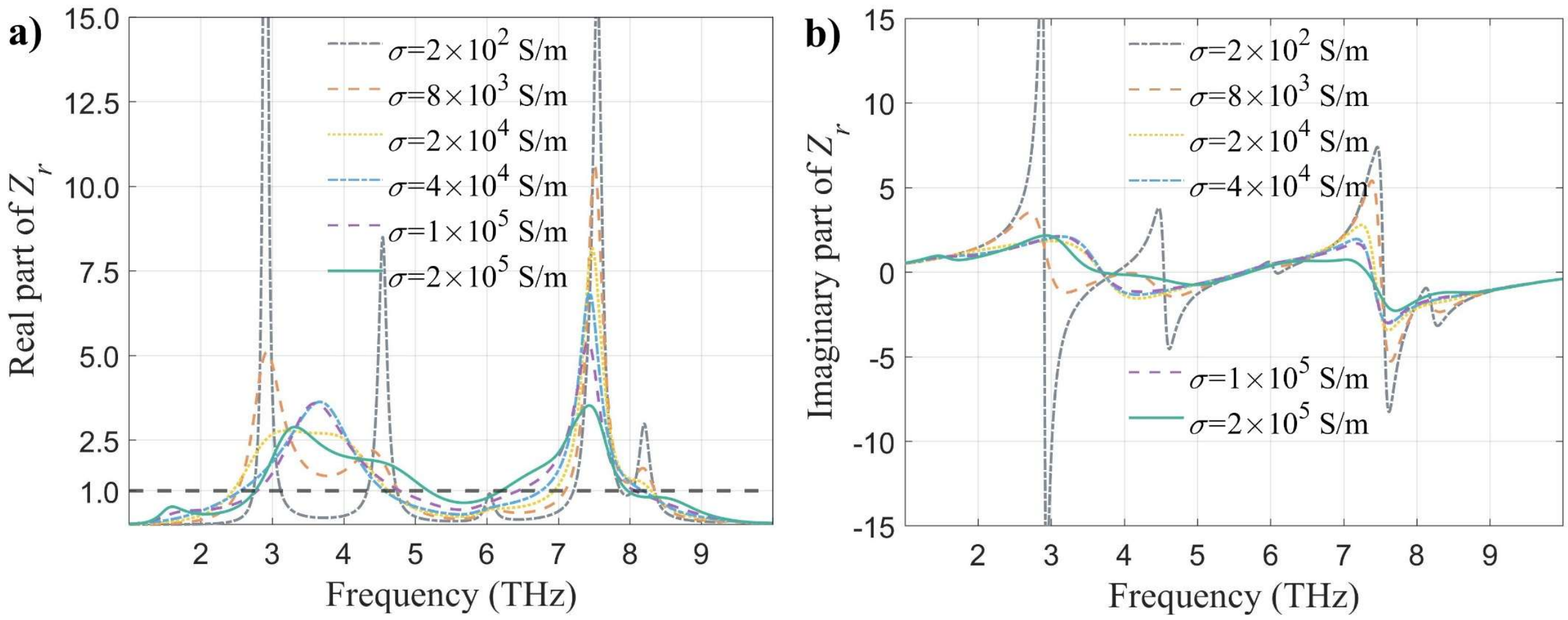

Fig. 9. a) The real and b) the imaginary parts of the relative impedance ($Z_r$) for different conductivities of $VO_2$ under normal incidence. The optimized geometrical parameters are $P$=40 µm, $L_1$=20 µm, $L_2$=5 µm, $L_3$=4.5 µm, $L_4$=6 µm, $w$=2.5 µm, $t_s$=8 µm, and $t_{Au}=t_{VO_2}=200$ nm.

The electric current intensity and its direction at the interface of the hybrid metal-$VO_2$ resonator and the polyimide spacer for the normally incident wave at the different frequencies and different conductivities of $VO_2$ are shown in Fig. 10. The frequencies of 3.97, 6.05, and 8.11 THz correspond to the absorption peaks of the metasurface for $\sigma=2\times10^2$ S/m. At different frequencies and conductivities, the current intensity and its direction in the gold and $VO_2$ strips are different. When the conductivity of the $VO_2$ is $\sigma=2\times10^2$ S/m, the electric current flows with high intensity in the gold strips. However, as the conductivity of the $VO_2$ increases, the electric current induced in the $VO_2$ strips increases. The current distributions for $\sigma=1\times10^5$ and $2\times10^5$ S/m are almost the same and as can be seen in Fig. 8 their absorption spectra are also similar. For brevity, the current density on the ground-plane is not shown in Fig. 10. The current distribution on the ground-plane is antiparallel to that at the interface of the resonator and the spacer, therefore, a current loop is formed. Consequently, magnetic resonance also plays a role in the absorption of the metasurface.

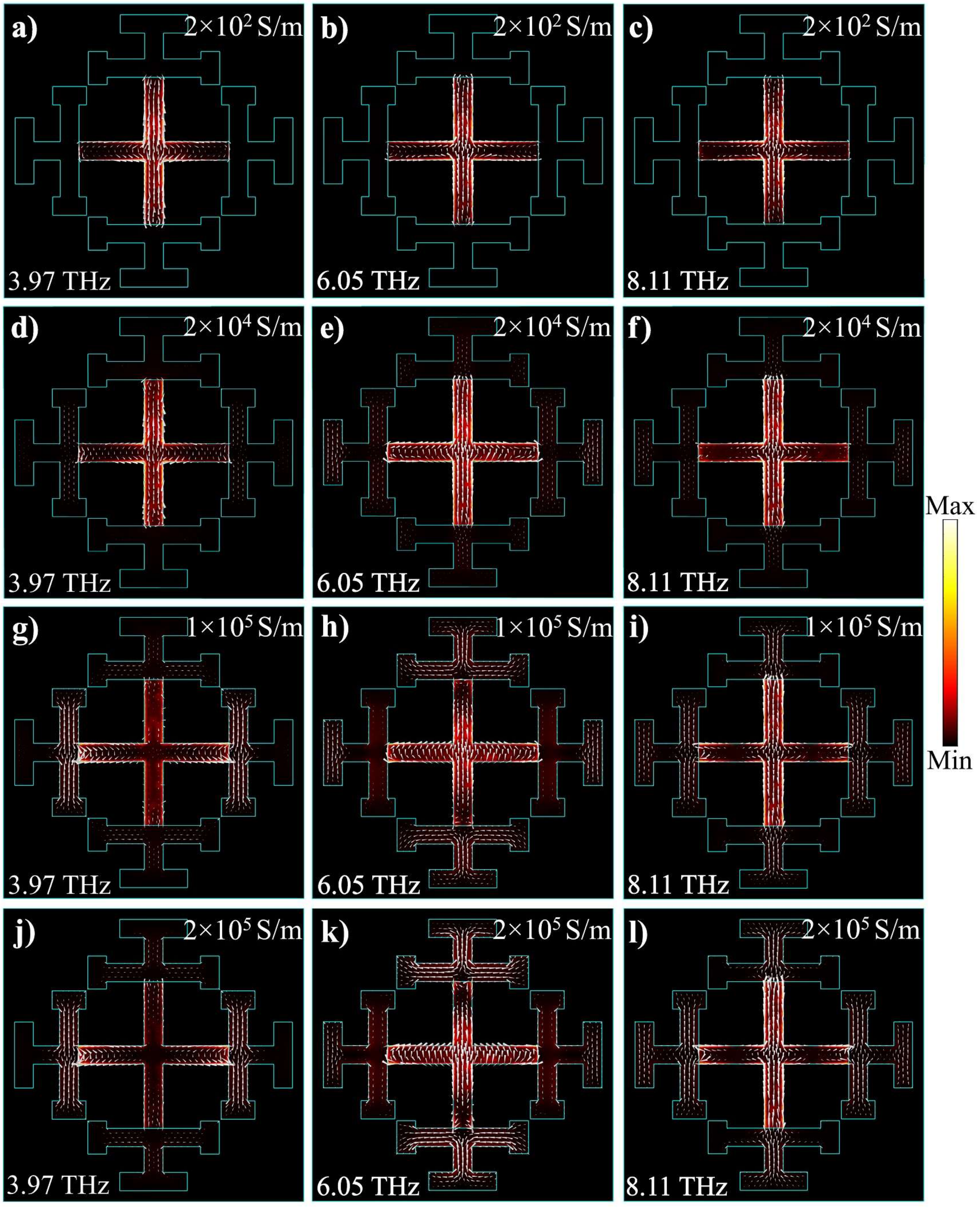

Fig. 10. The intensity and direction of current at the interface of the hybrid resonator and the dielectric spacer for different conductivities of $VO_2$ and at different frequencies under normal incidence. The logarithmic scale is used for the cones showing the current direction.

### 3.3 Angle-dependence of the absorption

Ideally, the metasurfaces should be insensitive to the incident angle. To evaluate the performance of the designed metasurface, the incident angle ($\theta$) was varied from 0° to 85° with a step of 5° in the full-wave simulations. The absorption of the metasurface for the insulator and metallic phases of $VO_2$ is plotted in Fig. 11. In the insulator phase, the absorber is relatively sensitive to the incident angle as can be seen in Fig. 11(a). For normal incidence, there are three absorption peaks at frequencies of 3.97, 6.05, and 8.11 THz with absorption levels of about 0.61, 0.99, and 0.83. As the incident angle approaches 85°, the number of absorption peaks increases and at the frequencies of 3.71, 4.63, 5.96, 7.06, and 8.25 THz the absorption levels approach 0.48, 0.94, 0.74, 0.98, and 0.62, respectively. On the other hand, in the metallic phase, the absorber is relatively insensitive to the incident angle (Fig. 11(b)). As the incident angle increases, the absorption bandwidth gets narrower. However, for incident angles up to 65° the absorption remains above 0.5 in the 3.57-8.45 THz frequency range.

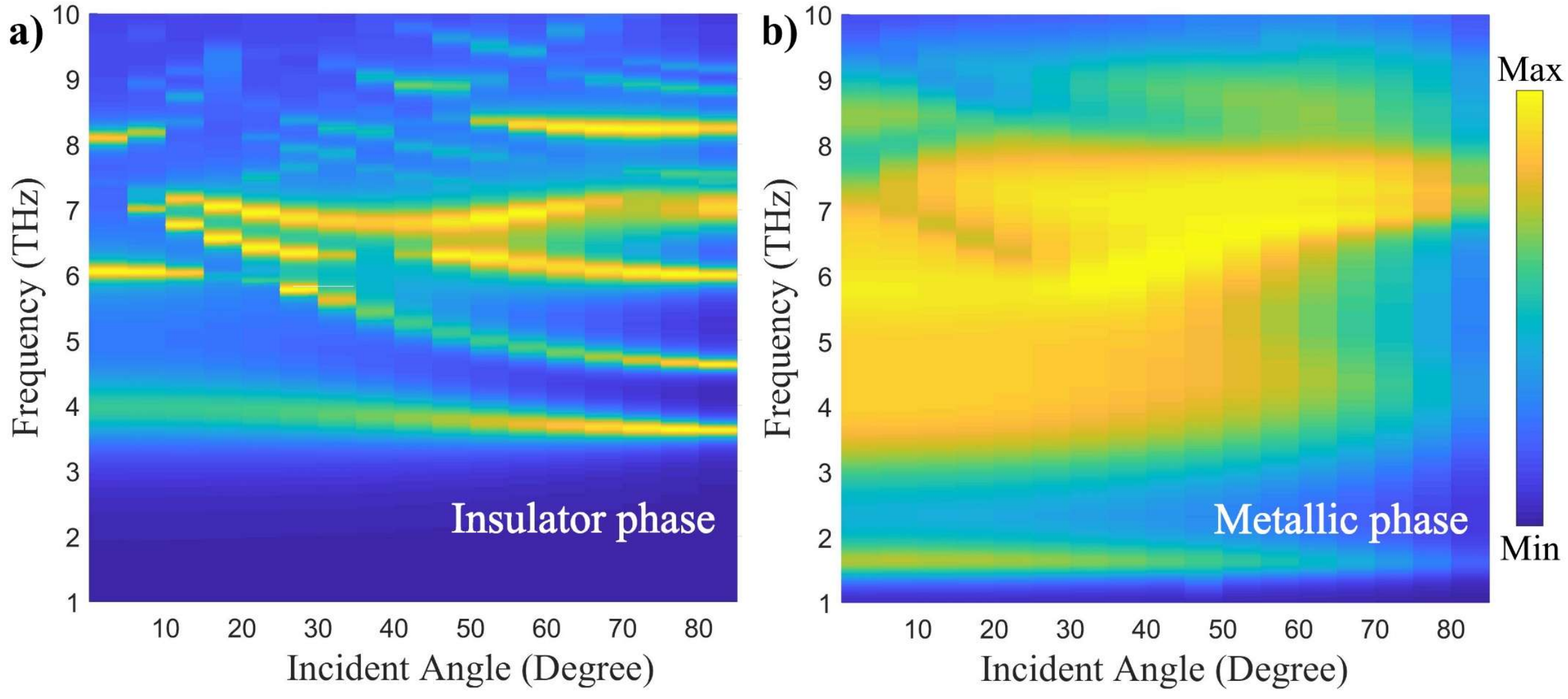

Fig. 11. Absorption spectra of the absorber vs the incident angle for a) insulator and b) metallic phases of $VO_2$. The optimized geometrical parameters are $P$=40 µm, $L_1$=20 µm, $L_2$=5 µm, $L_3$=4.5 µm, $L_4$=6 µm, $w$=2.5 µm, $t_s$=8 µm, and $t_{Au} = t_{VO_2} = 200$ nm.

Finally, we compare the performance of the designed narrowband-to-broadband switchable absorber with previous studies in Table 1. In this table, we compare the method of the bandwidth switching, the number of layers in the absorber, the performance of the absorber in the narrowband and broadband states, and insensitivity to polarization. Here, we define the fractional bandwidth as $BW_F=100\times BW/f_c$ where BW is the half-power bandwidth and $f_c$ is the center frequency. The switchable absorber [61] is composed of six layers and its fractional bandwidth is 15.6 and 100% in the narrowband and broadband state of operation, respectively. In [62], a bidirectional absorber based on a multilayer stack of metal-dielectric with six layers has been proposed where its bandwidth is narrow or broad depending on the direction of the incidence. This limits its application as a narrowband-to-broadband switchable absorber. A switchable absorber has been designed based on a six layer structure consisting $VO_2$ and graphene with the fractional bandwidth of 18.9 and 103% for its narrowband and broadband states, respectively [63]. The use of graphene complicates its fabrication, however, it offers electrical tunability of center frequency and absorption level. Nanograting structure composed of a $VO_2$ layer sandwiched between Si and a continuous Ag bottom reflector has been proposed as an actively tunable absorber [40]. In its narrowband absorption state, it has two absorption peaks with fractional bandwidths of 3.8 and 3.7% while its broadband absorption state offers only a fractional bandwidth of 17%. Our design compared to these switchable absorbers offers a relatively narrow bandwidth in its narrowband state while it offers a broader bandwidth in its broadband state. Moreover, the proposed absorber is composed of only three layers while the other absorbers have a higher number of layers complicating the fabrication process. We also discuss the performance limitations of the proposed absorber. In the insulator phase, there are three absorption peaks. For sensing applications, absorption peaks with higher quality factors are desired. However, the quality factor of the designed absorber at the main absorption peak is $Q$=6.05 THz/0.41 THz=14.75 which is low. In future works, an absorber with a higher quality factor should be pursued for applications such as sensors. Moreover, an absorber with less sensitivity to the incident angle can also be pursued in future works.

Table 1. Comparison of narrowband-to-broadband switchable absorbers.

| Ref. | Method | No. of Layers | Narrowband performance | Broadband performance | Polarization insensitive |
|---|---|---|---|---|---|
| [61] | PCM | 6 | 0.74 absorption @ 0.67 THz with $BW_F$≈15.6% | $BW_F$≈100% $f_c$=0.63 THz | Yes |
| [62] | Changing the direction of incidence | 6 | 1.0 absorption @ 388.8 THz with $BW_F$≈2.6% | $BW_F$≈97% $f_c$=307.5 THz | Yes |
| [63] | PCM and Graphene | 6 | 1.0 absorption @ 1.37 THz with $BW_F$≈18.9% | $BW_F$≈100% $f_c$=1.68 THz | Yes |
| [40] | PCM | 4 | 0.97 and 1.0 absorption @ 224.7 and 273.9 THz with $BW_F$≈ 3.8 and 3.7% | $BW_F$≈17% $f_c$=247.4 THz | No |
| This work | PCM | 3 | 0.99 absorption @ 6.05 THz with $BW_F$≈5.8% | $BW_F$≈102% $f_c$=5.68 THz | Yes |

## 4. Conclusion

In conclusion, we propose a polarization-insensitive terahertz metasurface absorber with reconfigurable bandwidth based on $VO_2$. Reversible phase-transition of $VO_2$ is exploited to switch from a narrowband absorber to a broadband absorber by tuning the working temperature. The metasurface is composed of a metal-$VO_2$ hybrid resonator and a lossy spacer placed on a metallic ground-plane. The resonator is a golden first-order cross fractal connected to $VO_2$ strips converting it to a third-order cross fractal. At room temperature, the designed metasurface efficiently absorbs in narrowband bandwidths with absorption exceeding 0.5 at the resonant frequencies of 3.97, 6.05, and 8.11 THz with 0.38, 0.35, and 0.22 THz bandwidths, respectively. At sufficiently high temperatures, $VO_2$ transforms into its fully metallic phase and, consequently, the metasurface switches to a broadband absorber with the FWHM of 6.17 THz centered around 5.79 THz. Due to the symmetry of the designed resonator, the absorber is insensitive to polarization and the broadband performance of the metasurface with absorption exceeding 0.5 is sustained over a wide range of incident angles up to 65° in the 3.57-8.45 THz frequency range.

## References

[1] M. Kadic, G. W. Milton, M. van Hecke, and M. Wegener, "3D metamaterials," *Nature Reviews Physics,* vol. 1, no. 3, pp. 198-210, 2019.

[2] Y. Liu and X. Zhang, "Metamaterials: a new frontier of science and technology," *Chemical Society Reviews,* vol. 40, no. 5, pp. 2494-2507, 2011.

[3] S. H. Badri and M. Gilarlue, "Silicon nitride waveguide devices based on gradient-index lenses implemented by subwavelength silicon grating metamaterials," *Applied Optics,* vol. 59, no. 17, pp. 5269-5275, 2020.

[4] L. La Spada and L. Vegni, "Near-zero-index wires," *Optics express,* vol. 25, no. 20, pp. 23699-23708, 2017.

[5] N. M. Estakhri, B. Edwards, and N. Engheta, "Inverse-designed metastructures that solve equations," *Science,* vol. 363, no. 6433, pp. 1333-1338, 2019.

[6] L. La Spada, C. Spooner, S. Haq, and Y. Hao, "Curvilinear metasurfaces for surface wave manipulation," *Scientific reports,* vol. 9, no. 1, pp. 1-10, 2019.

[7] N. J. Greybush, V. Pacheco-Peña, N. Engheta, C. B. Murray, and C. R. Kagan, "Plasmonic optical and chiroptical response of self-assembled Au nanorod equilateral trimers," *ACS nano,* vol. 13, no. 2, pp. 1617-1624, 2019.

[8] I.-H. Lee, D. Yoo, P. Avouris, T. Low, and S.-H. Oh, "Graphene acoustic plasmon resonator for ultrasensitive infrared spectroscopy," *Nature nanotechnology,* vol. 14, no. 4, pp. 313-319, 2019.

[9] Z. Lalegani, S. S. Ebrahimi, B. Hamawandi, L. La Spada, and M. Toprak, "Modeling, design, and synthesis of gram-scale monodispersed silver nanoparticles using microwave-assisted polyol process for metamaterial applications," *Optical Materials,* vol. 108, p. 110381, 2020.

[10] S. H. Badri and M. Gilarlue, "Coupling Si3N4 waveguide to SOI waveguide using transformation optics," *Optics Communications,* vol. 460, p. 125089, 2020.

[11] S. So and J. Rho, "Geometrically flat hyperlens designed by transformation optics," *Journal of Physics D: Applied Physics,* vol. 52, no. 19, p. 194003, 2019.

[12] S. H. Badri, H. R. Saghai, and H. Soofi, "Polygonal Maxwell's fisheye lens via transformation optics as multimode waveguide crossing," *Journal of Optics,* vol. 21, no. 6, p. 065102, 2019.

[13] S. H. Badri and M. M. Gilarlue, "Ultrashort waveguide tapers based on Luneburg lens," *Journal of Optics,* vol. 21, no. 12, p. 125802, 2019.

[14] F. Ding, S. Zhong, and S. I. Bozhevolnyi, "Vanadium dioxide integrated metasurfaces with switchable functionalities at terahertz frequencies," *Advanced Optical Materials,* vol. 6, no. 9, p. 1701204, 2018.

[15] G. Zheng, H. Mühlenbernd, M. Kenney, G. Li, T. Zentgraf, and S. Zhang, "Metasurface holograms reaching 80% efficiency," *Nature nanotechnology,* vol. 10, no. 4, pp. 308-312, 2015.

[16] L. Li *et al.*, "Electromagnetic reprogrammable coding-metasurface holograms," *Nature communications,* vol. 8, no. 1, pp. 1-7, 2017.

[17] W. Liu *et al.*, "Realization of broadband cross-polarization conversion in transmission mode in the terahertz region using a single-layer metasurface," *Optics letters,* vol. 40, no. 13, pp. 3185-3188, 2015.

[18] Y. Jiang, L. Wang, J. Wang, C. N. Akwuruoha, and W. Cao, "Ultra-wideband high-efficiency reflective linear-to-circular polarization converter based on metasurface at terahertz frequencies," *Optics express,* vol. 25, no. 22, pp. 27616-27623, 2017.

[19] M. El Badawe, T. S. Almoneef, and O. M. Ramahi, "A true metasurface antenna," *Scientific reports,* vol. 6, no. 1, pp. 1-8, 2016.

[20] M. Faenzi *et al.*, "Metasurface antennas: new models, applications and realizations," *Scientific reports,* vol. 9, no. 1, pp. 1-14, 2019.

[21] Q. Wang *et al.*, "A broadband metasurface-based terahertz flat-lens array," *Advanced Optical Materials,* vol. 3, no. 6, pp. 779-785, 2015.

[22] D. Jia *et al.*, "Transmissive terahertz metalens with full phase control based on a dielectric metasurface," *Optics letters,* vol. 42, no. 21, pp. 4494-4497, 2017.

[23] J. Wu *et al.*, "Liquid crystal programmable metasurface for terahertz beam steering," *Applied Physics Letters,* vol. 116, no. 13, p. 131104, 2020.

[24] M. Wei *et al.*, "Ultrathin metasurface-based carpet cloak for terahertz wave," *Optics express,* vol. 25, no. 14, pp. 15635-15642, 2017.

[25] Z. Liu *et al.*, "Dual-mode on-to-off modulation of plasmon-induced transparency and coupling effect in patterned graphene-based terahertz metasurface," *Nanoscale research letters,* vol. 15, no. 1, pp. 1-9, 2020.

[26] X. Liu, K. Fan, I. V. Shadrivov, and W. J. Padilla, "Experimental realization of a terahertz all-dielectric metasurface absorber," *Optics express,* vol. 25, no. 1, pp. 191-201, 2017.

[27] S. H. Badri and S. G. Farkoush, "Subwavelength grating waveguide filter based on cladding modulation with a phase-change material grating," *Applied Optics,* vol. 60, no. 10, pp. 2803-2810, 2021.

[28] S. H. Badri, M. M. Gilarlue, S. G. Farkoush, and S.-B. Rhee, "Reconfigurable bandpass optical filters based on subwavelength grating waveguides with a Ge2Sb2Te5 cavity," *J. Opt. Soc. Am. B,* vol. 38, no. 4, pp. 1283-1289, 2021.

[29] H.-S. P. Wong *et al.*, "Phase change memory," *Proceedings of the IEEE,* vol. 98, no. 12, pp. 2201-2227, 2010.

[30] G. W. Burr *et al.*, "Recent progress in phase-change memory technology," *IEEE Journal on Emerging and Selected Topics in Circuits and Systems,* vol. 6, no. 2, pp. 146-162, 2016.

[31] Y. Qu, Q. Li, K. Du, L. Cai, J. Lu, and M. Qiu, "Dynamic Thermal Emission Control Based on Ultrathin Plasmonic Metamaterials Including Phase-Changing Material GST," *Laser & Photonics Reviews,* vol. 11, no. 5, p. 1700091, 2017.

[32] C. Zhou, X. Qu, S. Xiao, and M. Fan, "Imaging Through a Fano-Resonant Dielectric Metasurface Governed by Quasi--bound States in the Continuum," *Physical Review Applied,* vol. 14, no. 4, p. 044009, 2020.

[33] C. Zhou *et al.*, "Optical radiation manipulation of Si-Ge2Sb2Te5 hybrid metasurfaces," *Optics express,* vol. 28, no. 7, pp. 9690-9701, 2020.

[34] F. Ding, Y. Yang, and S. I. Bozhevolnyi, "Dynamic metasurfaces using phase-change chalcogenides," *Advanced Optical Materials,* vol. 7, no. 14, p. 1801709, 2019.

[35] D. Wang, S. Sun, Z. Feng, and W. Tan, "Enabling switchable and multifunctional terahertz metasurfaces with phase-change material," *Optical Materials Express,* vol. 10, no. 9, pp. 2054-2065, 2020.

[36] Z. Song and J. Zhang, "Achieving broadband absorption and polarization conversion with a vanadium dioxide metasurface in the same terahertz frequencies," *Optics express,* vol. 28, no. 8, pp. 12487-12497, 2020.
[37] D. Yan, M. Meng, J. Li, J. Li, and X. Li, "Vanadium dioxide-assisted broadband absorption and linear-to-circular polarization conversion based on a single metasurface design for the terahertz wave," *Optics Express,* vol. 28, no. 20, pp. 29843-29854, 2020.
[38] J. Li *et al.*, "All-Optical Switchable Vanadium Dioxide Integrated Coding Metasurfaces for Wavefront and Polarization Manipulation of Terahertz Beams," *Advanced Theory and Simulations,* vol. 3, no. 1, p. 1900183, 2020.
[39] J. Huang, J. Li, Y. Yang, J. Li, Y. Zhang, and J. Yao, "Active controllable dual broadband terahertz absorber based on hybrid metamaterials with vanadium dioxide," *Optics express,* vol. 28, no. 5, pp. 7018-7027, 2020.
[40] A. K. Osgouei, H. Hajian, B. Khalichi, A. E. Serebryannikov, A. Ghobadi, and E. Ozbay, "Active Tuning from Narrowband to Broadband Absorbers Using a Sub-wavelength VO2 Embedded Layer," *Plasmonics,* pp. 1-9, 2021.
[41] S. Wang, L. Kang, and D. H. Werner, "Hybrid resonators and highly tunable terahertz metamaterials enabled by vanadium dioxide (VO 2)," *Scientific reports,* vol. 7, no. 1, pp. 1-8, 2017.
[42] T. Li *et al.*, "Terahertz bandstop-to-bandpass converter based on VO2 hybrid metasurface," *Journal of Physics D: Applied Physics,* 2021.
[43] J. Luo, X. Shi, X. Luo, F. Hu, and G. Li, "Broadband switchable terahertz half-/quarter-wave plate based on metal-VO 2 metamaterials," *Optics Express,* vol. 28, no. 21, pp. 30861-30870, 2020.
[44] S. Liu, H. Chen, and T. J. Cui, "A broadband terahertz absorber using multi-layer stacked bars," *Applied Physics Letters,* vol. 106, no. 15, p. 151601, 2015.
[45] Y. Liu, Y. Qian, F. Hu, M. Jiang, and L. Zhang, "A dynamically adjustable broadband terahertz absorber based on a vanadium dioxide hybrid metamaterial," *Results in Physics,* vol. 19, p. 103384, 2020.
[46] A. Zdagkas *et al.*, "Observation of toroidal pulses of light," *arXiv preprint arXiv:2102.03636,* 2021.
[47] N. Papasimakis, V. Fedotov, V. Savinov, T. Raybould, and N. Zheludev, "Electromagnetic toroidal excitations in matter and free space," *Nature materials,* vol. 15, no. 3, pp. 263-271, 2016.
[48] R. Xu, X. Liu, and Y.-S. Lin, "Tunable ultra-narrowband terahertz perfect absorber by using metal-insulator-metal microstructures," *Results in Physics,* vol. 13, p. 102176, 2019.
[49] J. Zhu *et al.*, "Ultra-broadband terahertz metamaterial absorber," *Applied Physics Letters,* vol. 105, no. 2, p. 021102, 2014.
[50] C. M. Watts, X. Liu, and W. J. Padilla, "Metamaterial electromagnetic wave absorbers," *Advanced materials,* vol. 24, no. 23, pp. OP98-OP120, 2012.
[51] H. Tao *et al.*, "A dual band terahertz metamaterial absorber," *Journal of physics D: Applied physics,* vol. 43, no. 22, p. 225102, 2010.
[52] Z. Song, M. Wei, Z. Wang, G. Cai, Y. Liu, and Y. Zhou, "Terahertz absorber with reconfigurable bandwidth based on isotropic vanadium dioxide metasurfaces," *IEEE Photonics Journal,* vol. 11, no. 2, pp. 1-7, 2019.
[53] G. Zhou *et al.*, "Broadband and high modulation-depth THz modulator using low bias controlled VO 2-integrated metasurface," *Optics express,* vol. 25, no. 15, pp. 17322-17328, 2017.
[54] M. R. M. Hashemi, S.-H. Yang, T. Wang, N. Sepúlveda, and M. Jarrahi, "Electronically-controlled beam-steering through vanadium dioxide metasurfaces," *Scientific reports,* vol. 6, no. 1, pp. 1-8, 2016.
[55] T. Wang *et al.*, "Thermally switchable terahertz wavefront metasurface modulators based on the insulator-to-metal transition of vanadium dioxide," *Optics express,* vol. 27, no. 15, pp. 20347-20357, 2019.

[56] D. Wang *et al.*, "Switchable ultrathin quarter-wave plate in terahertz using active phase-change metasurface," *Scientific reports,* vol. 5, no. 1, pp. 1-9, 2015.

[57] F. Miyamaru *et al.*, "Terahertz electric response of fractal metamaterial structures," *Physical Review B,* vol. 77, no. 4, p. 045124, 2008.

[58] M. Kenney, J. Grant, Y. D. Shah, I. Escorcia-Carranza, M. Humphreys, and D. R. Cumming, "Octave-spanning broadband absorption of terahertz light using metasurface fractal-cross absorbers," *Acs Photonics,* vol. 4, no. 10, pp. 2604-2612, 2017.

[59] J. D. Frame, N. G. Green, and X. Fang, "Modified Maxwell Garnett model for hysteresis in phase change materials," *Optical Materials Express,* vol. 8, no. 7, pp. 1988-1996, 2018.

[60] D. Smith, D. Vier, T. Koschny, and C. Soukoulis, "Electromagnetic parameter retrieval from inhomogeneous metamaterials," *Physical review E,* vol. 71, no. 3, p. 036617, 2005.

[61] Z. Song, A. Chen, and J. Zhang, "Terahertz switching between broadband absorption and narrowband absorption," *Optics express,* vol. 28, no. 2, pp. 2037-2044, 2020.

[62] F. Wang *et al.*, "Bidirectional band-switchable nano-film absorber from narrowband to broadband," *Optics Express,* vol. 29, no. 4, pp. 5110-5120, 2021.

[63] M. Zhang and Z. Song, "Terahertz bifunctional absorber based on a graphene-spacer-vanadium dioxide-spacer-metal configuration," *Optics express,* vol. 28, no. 8, pp. 11780-11788, 2020.